# A Novel Hybrid Biometric Electronic Voting System: Integrating Finger Print and Face Recognition

SYED SHAHRAM NAJAM*, AAMIR ZEB SHAIKH*, AND SHABBAR NAQVI**

## ABSTRACT

A novel hybrid design based electronic voting system is proposed, implemented and analyzed. The proposed system uses two voter verification techniques to give better results in comparison to single identification based systems. Finger print and facial recognition based methods are used for voter identification. Cross verification of a voter during an election process provides better accuracy than single parameter identification method. The facial recognition system uses Viola-Jones algorithm along with rectangular Haar feature selection method for detection and extraction of features to develop a biometric template and for feature extraction during the voting process. Cascaded machine learning based classifiers are used for comparing the features for identity verification using GPCA (Generalized Principle Component Analysis) and K-NN (K-Nearest Neighbor). It is accomplished through comparing the Eigen-vectors of the extracted features with the biometric template pre-stored in the election regulatory body database. The results of the proposed system show that the proposed cascaded design based system performs better than the systems using other classifiers or separate schemes i.e. facial or finger print based schemes. The proposed system will be highly useful for real time applications due to the reason that it has 91% accuracy under nominal light in terms of facial recognition.

## 1. INTRODUCTION

Electoral Systems empower the citizens of a country to elect parliament members of their choice. Paper based electoral system is a classical method to accomplish the said task. In this method, printed votes are submitted to various election booths of country at least one day before the election. After the election timings, sealed boxes containing votes are opened in front of all the legitimate members of booth and counted. The information of counted votes is submitted to a centralized station along with bags of paper votes. The central station compiles and publishes the names of winners and losers through television and radio stations. This method is useful only if the whole process is completed in a transparent way. However, there are some drawbacks to this system. These include higher expenses, longer time to complete the voting process, fraudulent practices by the authorities administering elections as well as malpractices by the voters [1]. These challenges result in manipulated election results.

Corresponding Author (E-Mail: shahramnajam.neduet@gmail.com)
*       Department of Electronic Engineering, NED University of Engineering & Technology, Karachi
**      Department of Computer Systems Engineering, Baluchistan University of Engineering & Technology, Khuzdar.

*A Novel Hybrid Biometric Electronic Voting System: Integrating Finger Print and Face Recognition*

Electronic Voting Systems provide efficient and reliable technique to empower citizens of a country or members of an organization to select a person of their choice. These systems can be classified into supervised, hybrid and remote voting styles. Supervised voting also known as offline voting is typically administered by electoral organizations. In this scheme, voting machines are located at polling machines. However, these machines are not connected with a centralized system for cross-verification or any other purpose. Hybrid voting schemes are supervised by election organizing members, however, the machines are connected with internet, Remote voting refers to the schemes which are not administered by any supervising staff and the machines are connected with internet [2]. Benefits of using Biometrics in a voting system is to accurately recognize the voter which enables the election administrators to reduce the error rates by reducing fraudulent and bogus votes. Besides, it also results in cost efficiency, improving physical safety and increasing convenience to the users [3].

In this regard, various authors have developed the electronic voting systems. A smart card based voting system is developed by[4]. This smart card system has temporary and permanent storage facilities. To address fraudulent practices, this card also contains biometric information of the end user which can be authenticated by the system. Sehr [5] present a computerized voting system to address issues including low attendance of voters, higher administration and operation costs, longer time of tabulation, and inconvenience for voters, rigid voting guidelines, and inadequate security protection.

Tagawa[6], present an innovative electronic voting system. The proposed system encodes voting information. This system consists of voting unit, polling administration unit, voter list administration unit and ballot or counting unit. After the vote is caste, the information is sent to polling administration unit along with the smart card number in encrypted manner. During the comparison if the information is found to be doubtful the vote will be rejected. Otherwise it can be preceded to the ballot counting unit. It is an effective system with proper data encryption and secrecy but it lacks one feature i.e. multiple votes by a single user.

Evertz [7] presents a system using WAN (Wide Area Network) which is connected to a server at the election office containing the database of all the voters. First the voter has to verify its identity by facial recognition, in which features are extracted from the face of the voter and compared with pre-stored features in a database. Upon matching of the identity, a window will pop up on the screen of the computer where the voter can cast its vote. But the facial recognition system used and employed is quite in-effective having a success percentage of only 58% and a response time of 15 seconds. Besides, it lacks any data encryption or security for the secrecy of the ballot. Thus rendering it in-effective for use in real-time.

To improve the confidentiality and privacy of the electronic voting systems, most of the systems use Mixnet or homomorphic encryption techniques [8]. Additionally, authors also claim that the homomorphic encryption is more appropriate for the situation with several election candidates as well as elections with neutral votes. The electronic voting system is implemented extensively in developed countries such as USA. Awad and Leiss [9], present a comprehensive study of conventional and electronic voting systems in USA along with their disadvantages. Alomari and Irani [10], present e-voting for a developing country, hence they concluded that the factors that Influence the adoption of e-voting includes trust in internet, trust in government, attitudes, website design, and compatibility including many others. Pesado et. al. [11], have presented the challenges and solutions of electronic voting system preferably for Argentina. Additionally, they have presented the characteristics of three different voting types, these include on site electronic voting system, partially onsite and remote voting system. Furthermore, policy considerations are also provided for the implementation of the proposed system.

Jacobs and Oostdijk [12], present a system that uses bar coded identifiers which are assigned either randomly or pseudo randomly in the form of combination of numbers and alphabets. These encrypted codes provide security from any illegal intervention.Using different identifiers makes this system secure in comparison to others. The voter will then have to scan the bar-code and then the system will decode and compare the code assigned with that of the database. Upon a perfect match the voter will be allowed to vote. Awan et. al. [13], implement a fingerprint based electronic voting system using Raspberry Pi board. Vidyasree et. al. [14], fuse the fingerprint and facial data to improve the identification of a voter through multimodal system. The results show a reasonable amount of improvement in comparison to unimodal system. Das et. al. [15], store biometric information of the user i.e. fingerprints on RF ID tags for designing an improved electronic voting machine. The proposed system also integrates the GSM module to disseminate information from the local station to other stations.

We develop and present an electronic voting system to eradicate fraudulent practices during public elections by involving double user identification checks i.e. facial recognition and finger print based identification methods. Facial recognition is accomplished through a feature-extraction based machine learning algorithm, while finger print based identification is achieved through pattern recognition method. The facial recognition is accomplished through cascading of Global Principal Component Analysis and K nearest Neighbor algorithms. The proposed method will provide better accuracy in comparison to a single identification method.

The Rest of the paper is organized as follows: System model of the proposed setup is presented in Section II. Section III presents the performance evaluation of the proposed algorithm. Section IV presents the conclusion of the paper.

## 2. THE PROPOSED SYSTEM

In this section, a brief description of various hardware units is presented that are integrated in proposed project to achieve the improved results for the proposed electronic voting system, as shown in Fig. 1.

**Microcontroller:** A microcontroller can be defined as an integrated circuit that contains a core processor and memory [16]. Microcontroller is also known as an embedded system, capable of storing, processing and transferring data and information between various peripherals interfaced with it on some logic, i.e. like a coordinating body of a circuit. With the advancement in the field of electronic technology especially in microelectronics and embedded system development, various development boardsare available. These boards include Arduino-UNO, Texas Instruments MSP 430 Launchpad, Nanode, Pinguino PIC 32, Teensy 2.0, Raspberry Pi and many others. These boards not only provide microcontroller facility to the end user but also an interfacing capability to connect different devices i.e. Bluetooth, Zigbee, LAN and WLAN (Wireless LAN also called WiFi). The proposed system (in our research) uses Arduino-UNO board due to good processing speed as well as memory, and capable of interfacing, controlling and monitoring of data flow [17].

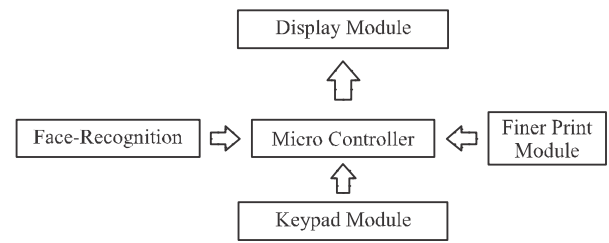

*FIG. 1. BLOCK DIAGRAM OF CR BASED BIOMETRIC ELECTRONIC VOTING SYSTEM*

**Fingerprint Module:** Unique finger impression recognition or fingerprint authentication indicates the mechanized strategy for checking a match between two human fingerprints [18]. The examination of fingerprints for coordinating purposes requires the correlation of components of the print design. The extracted parameters of a finger pattern include edges and minutia focuses [19]. These distinct features of a biological pattern give uniqueness to a human being.

The mechanized method for the verification of a fingerprint is done by using an electronic device called Fingerprint Verification Module, which captures the unique pattern of a fingerprint in the form of a computerized digital image. The digitally captured images are then processed to prepare a biometric template. This biometric layout is an accumulation of extricated elements which is stored and utilized for coordinating and matching [20]. The proposed system uses a finger print verification module developed by Future Electronics Egypt.

**Facial Recognition System:** Facial recognition system or facial acknowledgement framework is defined as an application capable of detecting and recognizing a person from a digitally processed image [21]. This unit comprises of facial recognition algorithms which includes facial detection, facial feature extraction, formation of biometric template by compression and formation of Eigen vectors and their comparison. Many popular facial recognition algorithms are available in literature that include PCA (Principal Component Analysis) using Eigen faces, LDA (Linear Discriminate Analysis), Fisher-face algorithm andDynamic link matching [22].

The proposed system incorporates the facial recognition algorithm, developed by [23]. The details of the algorithm and its working details are provided in the next section of this paper. A flow diagram of the proposed algorithmic setup is shown in Fig. 2. The input image is processed to be utilized by trained classifiers that produce a final decision of either recognized or unrecognized.

## 3. WORKING PROCEDURE

In this section, a brief working procedure of biometric data extraction and processing is presented. Fig. 3 shows the registration steps to be taken for the new voter registration into the proposed voting system. Fig. 3 shows the execution process of the proposed electronic voting system.

As shown in Fig. 3, the registration of the voter begins by the start of the counter for assigning a voter number to each voter. The message is displayed on the screen to place the face in front of the camera, the image is captured and normalized and divided into 24X24 sub-windows. Thus, distinct features are extracted and a vectored biometric layout of the facial image is formed. The resulting biometric template can be used to train the classifier using Adaboost trainer and then a codebook for the Eigen vectors is formed, then the generated biometric layout is saved in the database against the encrypted ID number. This ends the first step towards recognition of facial features. Then a message is displayed to place the thumb on the scanner, the thumb sensor scans and forms a biometric layout of the thumb of the voter and stores it against the same encrypted voter number in the database. Now both facial and thumb print biometric templates are compared with the pre-stored biometric layouts in the database in order to eradicate registration of the same voter multiple times. In case the registered voter tries to repeat the registration step again, the registration is rejected.

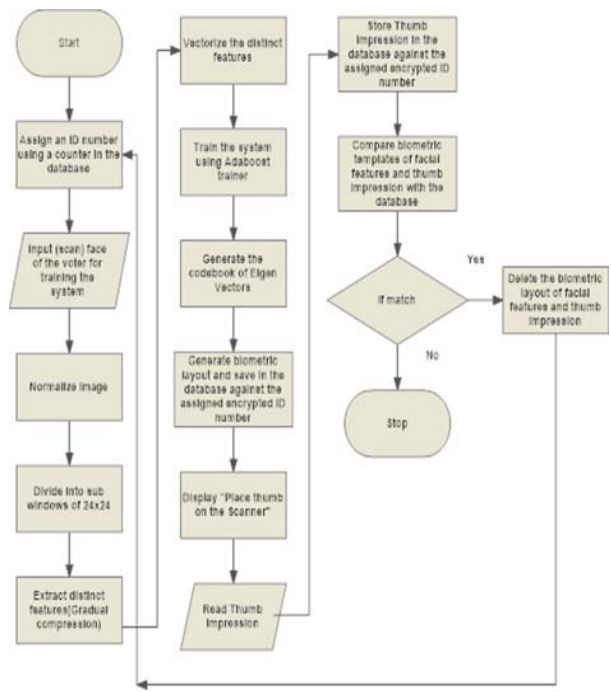

*FIG. 3. SHOWING THE REGISTRATION PROCESS OF VOTER*

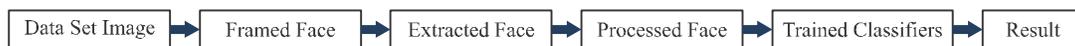

*FIG. 2. BLOCK DIAGRAM FOR FACE RECOGNITION*

During the voting process, a message is displayed to place the thumb on the thumb sensor/scanner; a biometric layout is generated and is compared with the database in order to find a match. In case the thumb impression is not found in the database, an error is displayed and a message is generated for the relevant users. In case a match is found, a message is displayed for the voter to place the face in front of the camera. The image is then normalized, 24X24 sub-windows are formed and features are extracted. The distinct features are vectored and are then compared with the biometric layout in the database. If a match is found, the voter is allowed to cast the vote. But in case no match is found, an error is displayed and a message is generated to the relevant authority, as shown in Fig. 4.

In this section, the detailed process of the individual steps is presented.

**Facial Recognition System:** The facial recognition system is the most significant feature of the proposed hybrid biometric electronic voting system. The algorithms used for facial recognition usually can be categorized into two methods firstly geometric which compare the geometry of distinct features and analyze the relative position, size and shape of ears, eyes nose and jawbones and secondly photometric which is a statistical methodology to distill a picture into statistical values and compares the values with the layout [24].

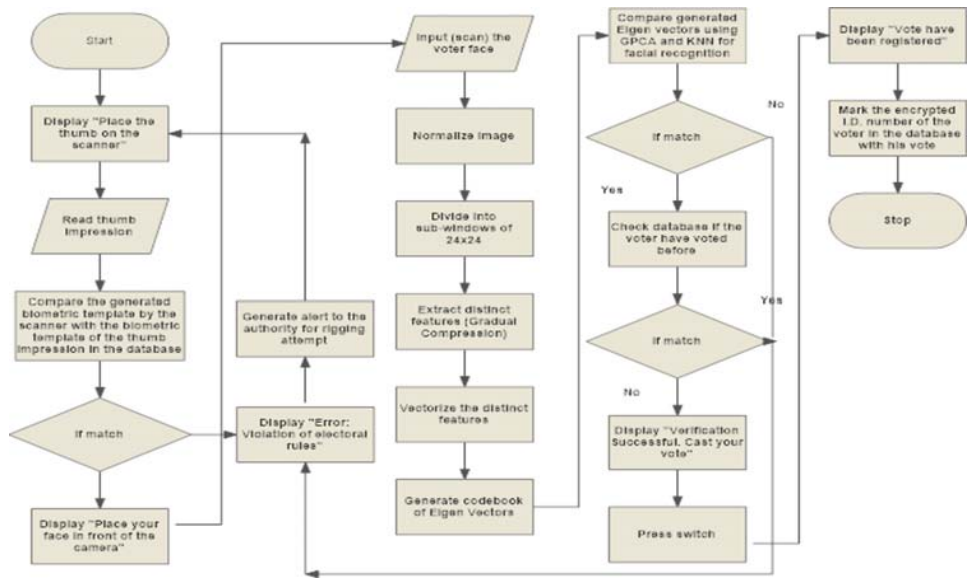

*FIG. 4. EXECUTION PROCESS FOR PROPOSED ELECTRONIC VOTING SYSTEM*

The algorithm used in the proposed system is based on the principle of feature extraction. Feature extraction in image processing may be defined as being a set of initial value derived from an object in the form of a pattern which is informative and useful for machine learning. The algorithm can be implemented using three steps i.e. Haar feature selection, creation of an integral image and Adaboost training [25].

The facial features are detected and analyzed using Haar feature selection like the positioning, distance and the geometric shape of the eyes, nose, ears and jaw bones and then using the information driven from Haar-Feature selection, an integral image is formed [26]. The process of face detection and feature selection using Haar feature selection can be referred in Fig. 5.

A sub-window of 24x24 pixels can exhibit a total of 162,336 possible features and it would be time consuming as well as expensive and considered to be quite an impractical approach for the facial recognition [27]. Hence Adaboost trainer is used which eliminates the scanning of all insignificant features and also train the classifiers to recognize the relevant features. Once an integral compressed biometric template of two-dimensional is formed, the features stored in the layout are converted into as set of Eigenvectors and thus an Eigen face is formed. The formation of Eigen face is to speed up the analysis and to reduce the response time as shown in Fig. 6.

Facial recognition is implemented through a cascaded classifier of GPCA and KNN algorithms. KNN is a non-parametric formula used in classification of data. It is also used in pattern recognition. It is one of the simplest algorithms of machine learning for pattern recognition [28]. PCAis an algorithm that convertsthecorrelated elements to linearly uncorrelated elements through orthogonal transformation. In Generalized PCA, the condition of orthogonally is removed to consider an arbitrary number of spaces of unknown and different dimensions [29].

The cascaded classifier uses the comparison of Eigenvectors of the stored bio-metric template with the digital image of the voter generated at the time of voting and then compares the nearest numbers of similarities by introducing a test vector from the live scan of the voter. If the similarities is less than 90% keeping in mind the environmental light and tolerated offset angles, the similarities will be rejected and the voter won't be able to cast his/her vote.

Although many methods of face detection are present of which the cascaded classifier method by using local PCA

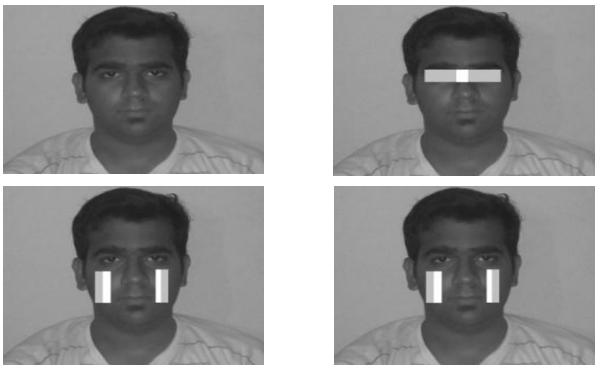

FIG. 5. FACE DETECTION USING VIOLA JONES ALGORITHM WITH RECTANGULAR HAAR FEATURE SELECTION

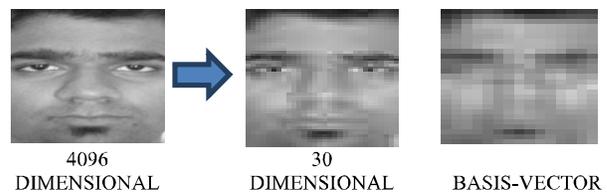

4096 DIMENSIONAL     30 DIMENSIONAL     BASIS-VECTOR

FIG. 6. FEATURE REDUCTION AND FORMATION OF EIGEN-FACE

and LDA but their resulting accuracy is quite low as compared to GPCA and KNN**. Fig. 7** shows the comparison of the accuracy of GPCA and KNN with LPCA and LDA in the next section of this paper.

Apart from that the response time for the cascaded classifier of GPCA and KNN method is quite fast and responsive as compared to the other methods usually employed for face-recognition which will be discussed in the Results and Discussion section of this paper.

**Fingerprint Processing:** The method employed by the finger print module is the optical method. Optical finger print verification technique maybe defined as the formation of a biometric template from the digitally computerized image for verification using visible light [30]. The surface for scanning the finger print is called as touch surface and underneath there is a light-transmitting phosphor layer which enlightens the surface of the finger. The light reflected from the finger goes through the phosphor layer to a variety of strong state pixels which captures a visual picture of the finger print [31].

The algorithm used for the finger print verification is a pattern-based verification in which the digital image of the finger print is compared with the previously stored bio-metric layout on the basis of similarities of the minutiae features like ridge ending, bifurcation, and short ridge [32]. The Fig. 8 shows the minutiae features of a finger.

The pre-stored template containing the features of minutiae features are compared with the finger print of the voter and if the comparison yields less than 90% comparison keeping in mind the tolerated offset angle, then the voter won't be allowed to vote.

## 4. RESULTS AND DISCUSSION

In this section, the results of the proposed electronic voting system are presented along with comparison with other systems.

The distinct and outstanding features of the proposed system are that the facial recognition algorithm is unique and the accuracy yielded as compared to other cascaded algorithms is high. The Table 1 shows the result of the experimental outcome of the electronic voting system for different K-Values:

The finger print module is also tested and its accuracy is shown in Table 2.

The Fig. 9 shows a comparison between KNN and a cascaded classifier i.e. GPCA and KNN. The results show that the cascaded system gives higher accuracy than individual KNN classifier.

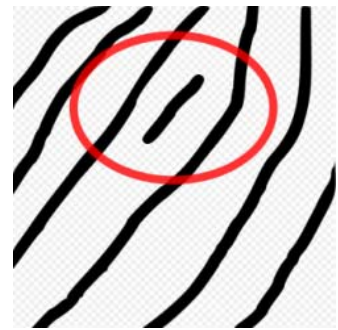

*FIG. 8. RIDGE DOT [30]*

**TABLE 1. FACIAL TESTING OUTCOME**

| Faces tested | Correct | Incorrect | Missed | Accuracy |
|---|---|---|---|---|
| 100 | 91 | 3 | 6 | 91% |

**TABLE 2. FINGER PRINT TESTING OUTCOME**

| Tested | Correct | Incorrect | Missed | Accuracy |
|---|---|---|---|---|
| 100 | 98 | 0 | 2 | 98% |

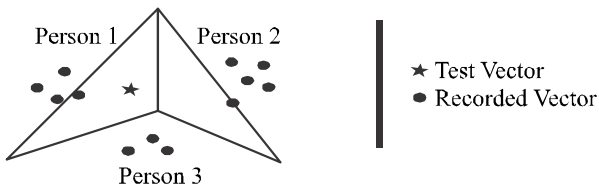

*FIG. 7. USING K-NN METHOD BY INTRODUCTION OF TEST VECTOR*

The use of cascaded classifier of KNN and GPCA rather than just using KNN and the outcome comparison of there accuracy with respect to changing number of K-values compared in a single cycle can be seen in Fig. 9 having an accuracy of 91% for a preset value of k=1 in the implemented system.

Also from the results and comparison of outcome of other studies and research papers, the accuracy of the outcome of separate and paired classifiers at a constant dimension is shown in Fig. 10 for comparison of 1764 distinct features and a K-value of 1 (for algorithms using K-NN).

the experiment carried out yielded the following results which are interpolated in Fig. 11 which shows a relation between distinct features in pixelated form stored in a biometric template and its effect on time response and was found to increase with the increasing number of distinct features and founded that the algorithm had a response time of 4.32 seconds for a 1764 distinct features and k-value=1 (k-value is the number of features compared per cycle) as preset in the system for real-time implementation.

Fig. 12 shows the comparison of the accuracy of GPCA and KNN with LPCA and LDA with respect to the distinct features compared with the features stored in the biometric template as preset in the algorithm for testing and having an accuracy of 91% (approximately) and k-value=1 (k-value is the number of features compared per cycle) for the 1764 distinct features.

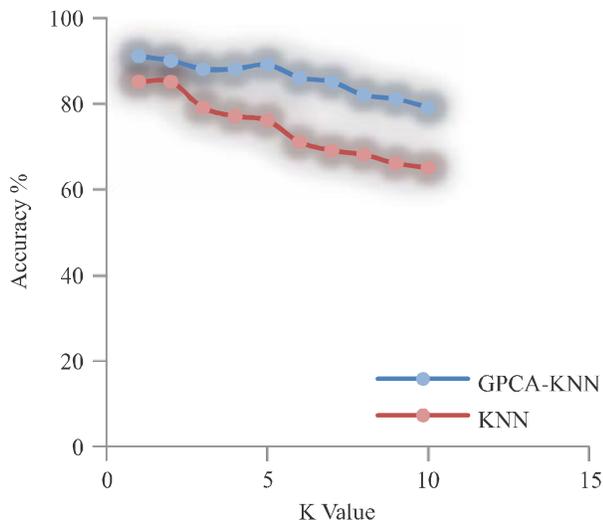

FIG. 9. COMPARISON OF ACCURACY BETWEEN KNN AND GPCA+KNN

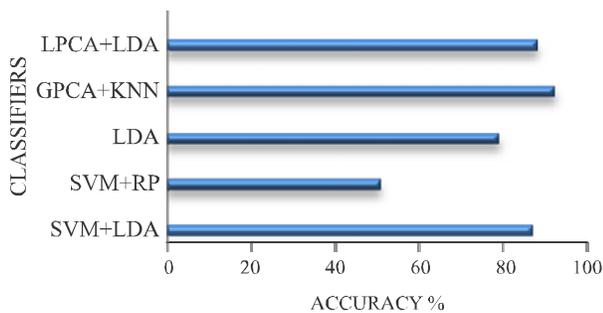

FIG. 10. ACCURACY OF DIFFERENT CLASSIFIERS

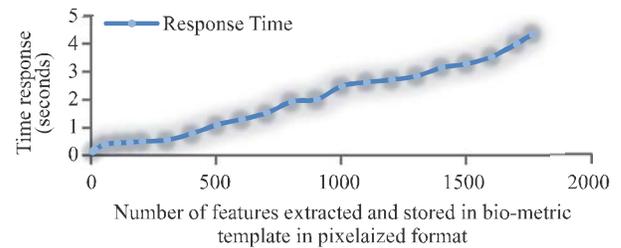

FIG. 11. RESPONSE TIME WITH VARYING DIMENSIONS

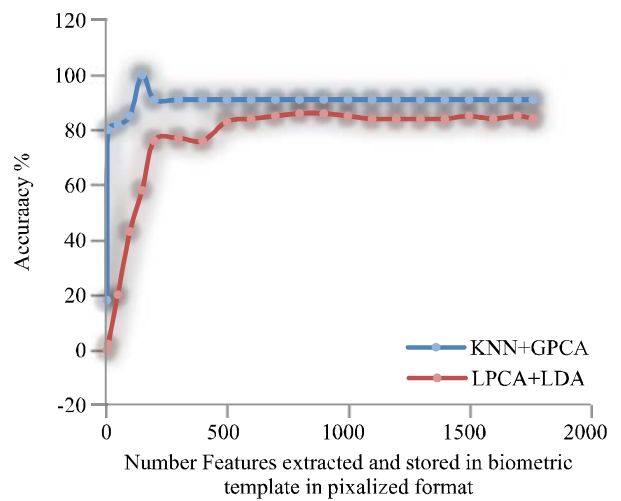

FIG. 12. COMPARISON OF ACCURACY WITH DIFFERENT CASCADED CLASSIFIERS

## 5. CONCLUSION

A novel hybrid biometric voting system is proposed, implemented and analyzed for fair polling process during general elections in developing countries including Pakistan, Nepal, Sri Lanka and others. The investigation results show 91% accuracy of the proposed system. With the implementation of GPCA and KNN cascaded classifiers that are discussed in the previous section of the paper. Additionally, the proposed system also involves the finger print based security feature to provide additional authenticity of the voter. The future work will be to incorporate security features in the proposed system by introducing encryption algorithms.

## ACKNOWLEDGMENT

The authors would like to thank the Administration of NED University of Engineering & Technology, Karachi, Pakistan, for providing resources to complete this research.